# A Critical Path Approach for Elucidating the Temperature Dependence of Granular Hopping Conduction

*Tsz Chun Wu*[1], *Juhn-Jong Lin*[2,3], *and Ping Sheng*[1,*]

[1]Department of Physics, Hong Kong University of Science and Technology, Clear Water Bay, Kowloon, Hong Kong

[2]Institute of Physics and Department of Electrophysics, National Chiao Tung University, Hsinchu 30010

[3]Center for Emergent Functional Matter Science, National Chiao Tung University, Hsinchu 30010

## Abstract

We revisit the classical problem of granular hopping conduction's $\sigma \propto \exp\left[-(T_o/T)^{1/2}\right]$ temperature dependence, where $\sigma$ denotes conductivity, $T$ is temperature, and $T_o$ is a sample-dependent constant. By using the hopping conduction formulation in conjunction with the incorporation of the random potential that has been shown to exist in insulator-conductor composites, it is demonstrated that the widely-observed temperature dependence of granular hopping conduction emerges very naturally through the immediate-neighbor critical-path argument. Here immediate-neighbor pairs are defined to be those where the line connecting the two grains does not cross, or by-pass, other grains; and critical-path argument denotes the derivation of sample conductance based on the geometric percolation condition that is marked by the critical conduction path in a random granular composite. Simulations based on the exact electrical network evaluation of finite-sample conductance show that the configuration-averaged results agree very well with those obtained by the immediate-neighbor critical-path method, and both show very good agreement with experimental data on hopping conduction in sputtered metal-insulator composite $Ag_x(SnO_2)_{1-x}$, where $x$ denotes the metal volume fraction. The present approach offers a relatively straightforward and simple explanation for the temperature behavior that was widely observed over diverse material systems, but which has remained a puzzle in spite of the various efforts for its explanation.

*Email: sheng@ust.hk

## I. Introduction

In diverse composite materials comprising small conducting grains dispersed in an insulating matrix, an almost ubiquitously-observed temperature dependence of the conductivity is given by $\sigma = \sigma_o \exp\left[-(T_o/T)^{1/2}\right]$, where $\sigma_o$ and $T_o$ are the sample-dependent constants, and $T$ denotes the temperature [1-9]. This temperature dependence is noted to differ from that for the three-dimensional variable-range-hopping (VRH) conduction $\sigma = \sigma_{VRH}^o \exp\left[-(T_{VRH}^o/T)^{1/4}\right]$ not only in the value of the exponent (1/2 vs. VRH's 1/4), but also in the implied underlying physical picture of the hopping sites [10, 11]. Here $T_{VRH}^o$ and $\sigma_{VRH}^o$ are the VRH conduction constants related to the particular sample. Whereas in VRH the sites are point-like impurities with random on-site potentials [10-14], in granular hopping (GH) the sites are conducting grains with a distribution of size $d$. In hopping conduction the conductance $\Gamma_{ij}$ between two sites $i$ and $j$ is given by the product of probabilities for thermal activation and tunneling [15-18]:

$$\Gamma_{ij} = \Gamma_o \exp(-E_{ij}/k_B T)\exp(-2\chi S_{ij}), \tag{1}$$

where $\Gamma_o$ is a constant for a given sample or hopping conduction network, $\chi = \xi^{-1}$ is the tunneling constant with $\xi$ being the decay length of the wavefunction in the insulating matrix, $S_{ij}$ is the tunneling distance between the two sites, $k_B$ denotes the Boltzmann constant, and $E_{ij} = (1/2)\left[|E_i - E_j| + |E_i - \mu_F| + |E_j - \mu_F|\right]$. Here $E_i$ and $E_j$ are the capacitive charging energies of grains $i$ and $j$, respectively; and $\mu_F$ denotes the chemical potential for the whole sample. Since the capacitive charging energies $E_i$ and $E_j$ are always positive and above $\mu_F$, it follows that by letting $\mu_F = 0$ in what follows, we have $E_{ij} = \max[E_i, E_j]$, which can be simply understood by the fact that if $E_i > E_j$, then the transition from $i$ to $j$ involves the activation probability $\exp(-E_i/k_B T)$ for the charge carrier being on site $i$ in the first place. But if the transition is from $j$ to $i$, then it involves the product of two activation probabilities $\exp(-E_j/k_B T)\exp[-(E_i - E_j)/k_B T]$, which again leads to $\exp(-E_i/k_B T)$. The same argument applies if $E_j > E_i$ so that the activation probability is given by $\exp(-E_j/k_B T)$ for transition either from $i$ to $j$, or from $j$ to $i$. In equilibrium and at absolute zero temperature, all grains are assumed to be electrically neutral.

A crucial element of GH lies in the fact that $E_i$ is the capacitance (charging) energy of grain $i$ and hence proportional to $1/d_i$ [1, 3], where $d_i$ denotes the size/diameter of grain $i$. Below we shall assume the particles to be roughly spherical in shape so that $E_i = 2e^2/\kappa(x)\eta_i d_i$, where $e$ is the electronic charge, $\kappa(x)$ is an effective dielectric constant of the composite whose value is a function of the average volume fraction $x$ of the conducting grains. A dimensionless random constant $\eta_i$ is introduced to account for the deviation from $\kappa(x)$ owing to local dielectric screening environment of grain $i$. The average of $\eta_i$ should be around 1. Since $\eta_i$ is always multiplied with $d_i$, its effect may be regarded as a renormalization of the grain diameter, i.e., $\eta_i d_i = d_i^{(e)}$, so that $E_i = 2e^2/\kappa(x)d_i^{(e)}$. The distribution of $d_i^{(e)}$ should differ quantitatively from the distribution of the physical grain size $d_i$, but can remain qualitatively similar.

In a granular medium the Fermi energies of the neutral grains need not be at exactly the same level, owing to the existence of local electrical potential variations arising from charge trapping centers, which naturally exist in the dielectric, as well as interfacial (between metal and insulator) states that can cause random potentials on the order of tens of millivolts. This point has been emphasized by Adkins [19, 20] based on the experimental results on the electrical field effects [21-24]. He proposed the existence of a random potential that has a distribution $f(E)$ which is flat between $\pm E_i/2$, and zero beyond that. That is, $f(E) = 1/2E_i$ for $|E| < E_i$ and zero otherwise. This flat distribution of the random potential is supported by the frequency-dependent tunneling experiment of Cavicchi and Silsbee [23], which also showed that such a random potential distribution is consistent with the zero-bias tunneling conductance suppression. Hence the Fermi levels of the grains can be shifted by the random potential up to $\pm E_i/2$.

Since $E_i \propto 1/d_i^{(e)}$, its distribution is generally assumed to follow a log-normal distribution $N(E_i)$ [19, 25]; the reason (for the log-normal distribution) is that the granular metals are usually fabricated through the co-sputtering or evaporation approach. And the grains are formed through surface diffusion on the substrate that results in aggregation, via a random coagulation process that can be mathematically represented by the multiplication of independent random variables with a Gaussian distribution. Hence the logarithm of the final grain size should be normally distributed. Given $N(E_i)$, the effect of the random potential can be accounted for by the convolution [19] of $N(E_i)$ with $f(E - E_i)$, i.e., the *charge carrier excitation energy distribution* is given by

$$D(E) = \int_0^\infty N(E_i) f(E - E_i) dE_i. \qquad (2)$$

We show below that the random potential, in conjunction with the charging energy, is responsible for granular hopping's temperature dependence. If we use the form of $f(E)$ given above, then the excitation energy distribution $D(E)$ is essentially flat up to the peak energy of the lognormal distribution, after which it drops to zero in a smooth manner.

It should be especially noted that the approach presented in this work does not require a broad distribution of grain sizes or the charging energy. In fact, the present approach works perfectly well in yielding the observed granular hopping's characteristic temperature dependence for fairly narrow size distributions, which are actually the case determined experimentally.

There are a number of theories attempting to explain the $\sigma = \sigma_o \exp\left[-(T_o/T)^{1/2}\right]$ temperature dependence of the GH conduction [1, 3, 26-28]. However, each suffers from either un-reasonable values of the parameters required to fit the data, or unreasonable assumption(s) about the detailed microstructure of the granular medium [2, 20, 29-31]. In particular, there is an otherwise very successful theory, based on the constancy of the relative volume fraction $x$ of the conducting grains, imposed upon *each* $\Gamma_{ij}$, an assumption (constancy of $x$) that can only be valid when averaged over the whole sample [1, 3, 32, 33]. That is, since $E_i = 2e^2/\kappa(x)d_i$ (the effect of the local dielectric screening environment was ignored in such an argument, i.e., $\eta_i$ was set to be equal to 1) and if we assume that $S_{ij}/d_i = \mu$ depends only on the metal volume fraction $x$, then it follows that we can write $\Gamma_{ij} = \Gamma_o \exp(-E_{ij}/k_BT)\exp(-2\chi S_{ij}) = \Gamma_o \exp\left[-\frac{2e^2\mu}{\kappa(x)S_{ij}k_BT} - 2\chi S_{ij}\right]$. The exponent, as a function of $S_{ij}$, now has a maximum the position of which is obtainable by taking the derivative with respect to $S_{ij}$ and setting the result to zero. By substituting the maximum value of $S_{ij}$, $(S_{ij})_{max}$, back into the exponent and taking that as the sample conductance, one immediately obtains the $\exp\left[-(T_o/T)^{1/2}\right]$ temperature dependence. However, since $S_{ij}/d_i = \mu$ *cannot be valid for each individual* $\Gamma_{ij}$, hence the validity of this whole approach has been called into question. In this context, we recall a similar situation in the early days of Mott's proposed scenario for the VRH [10], in which the constancy of the density of states $\rho$ of the impurity sites was used to relate the $E_{ij}$ and $S_{ij}$. The $\sigma = \sigma^o_{VRH} \exp\left[-(T^o_{VRH}/T)^{1/4}\right]$ form was obtained through exponent optimization, in a similar manner as described above, except now $\rho = \rho_o = \left(E_i S_{ij}^{\tilde{d}}\right)^{-1}$ implies an $\exp\left[-(T^o_{VRH}/T)^{1/(\tilde{d}+1)}\right]$ temperature variation, with the exponent value ¼ for sample dimensionality $\tilde{d}=3$ [10, 11]. A similar (valid) criticism can be leveled at such an argument, since the density of states is inherently a concept valid only when considered with respect to the whole sample, instead of being valid for each hopping site and its immediate neighboring environment [10, 11].

Another popular theory for granular hopping is that by Efros and Shklovskii (ES for short) [12], in which the $T^{-0.5}$-law is derived by considering VRH in the presence of a (soft) Coulomb gap. However, in a series of papers plus a review, Adkins had shown that the application of the ES theory to granular metals is not consistent with the reasonable values of the parameters required [2, 19, 34]. We would like to note, however, that the prediction of the ES theory has recently been observed in low-temperature conduction in antidot graphene [35] and oxygen deficient ZnO films [36], where its applicability is not in doubt. As applied to granular systems, however, the VRH required by the ES theory means that there is a necessity for a charge carrier on a granule to be transported to a distant granule (i.e., beyond the immediate neighbors) in a single step. In order for such a process to be feasible with some finite probability, it must be supplemented by the so-called co-tunneling process [37] in which the charge carrier would stopover at some intermediate neutral grain on its way to the destination granule. In order for the intermediate grain to remain neutral, co-tunneling requires that the intermediate grain either emit a charge simultaneous with the arrival of the tunneling charge carrier, or that the stopover is so short in duration that the charge on the intermediate grain is in a quantum mechanical virtual state, hence its charging energy can be discounted.

It is also to be noted that the $\exp\left[-(T_o/T)^{1/2}\right]$ has been demonstrated numerically from non-nearest neighbor hopping [38].

In contrast to the ES theory, in the present approach the granular hopping law is shown to be derivable with consideration of immediate-neighbor hopping only, which is more physical for the diverse material systems to which the GH behavior applies. In particular, no co-tunneling process is necessary since only hopping between immediate neighbors is required. Here the term "immediate neighbor" is defined to mean that the straight line connecting the centers of two immediate-neighbor grains, does not either cross, or pass by, a third grain or other grains. Due to surface or shape irregularities, the distance of closest approach between the surfaces of two immediate neighbor grains, i.e., tunneling distance, can vary within a range that may be specified by a distribution function. Hence even though the hopping is limited to immediate neighbors only, one can still have a (limited range) variable range hopping since the tunneling distances among the immediate neighbors can vary, in contrast to the "nearest neighbor" connotation which implies a fixed minimum separation.

It is the purpose of this work to show that the critical path method (CPM), first proposed by Ambegaokar, Halperin, and Langer [16], can be adapted to derive the $\sigma = \sigma_o \exp\left[-(T_o/T)^{1/2}\right]$ temperature dependence for GH conduction in the context of immediate-neighbor hopping. The argument is the same to that used to justify the $\sigma = \sigma_{VRH}^o \exp\left[-(T_{VRH}^o/T)^{1/4}\right]$ form for VRH, based on the constancy of the density of states $\rho$. However, here we use the CPM in the context of immediate-neighbor hopping, instead of the VRH in which hopping conduction to all sites is allowed. This difference implies a deviation from the traditional use of density of states in the CPM. Instead, we appeal directly to the distributions of the charge carrier excitation energy and the tunneling distance. This aspect is made explicit below in Section III.

## II. The critical path argument

Consider a system of random sites with $\Gamma_{ij}$ being the conductance between any two localized sites *i* and *j*. Pick a value of $\Gamma$ and consider the immediate-neighbor pair *ij* connected if $\Gamma_{ij} \geq \Gamma$ and disconnected otherwise. If $\Gamma$ is sufficiently large, then only a few $\Gamma_{ij}$'s can satisfy the connection criterion, and the connected sites can form only disjoint clusters. Now lower the value of $\Gamma$ and apply the connection rule again. More links will be added to the existing clusters. By lowering $\Gamma$ continuously, the average size of the connected clusters will increase in size until at $\Gamma = \Gamma_c$, the percolation conductance, an infinite network of connected sites is obtained. The fact that there must be such a $\Gamma_c$ is guaranteed by the phenomenon of percolation.

It is argued that $\Gamma_c$, with the immediate-neighbor pair conductance given by the form of Eq. (1), must be representative of the sample conductance up to a (non-exponential) factor [39]. The reason is that for $\Gamma_{ij} \geq \Gamma_c$, the network is not percolating, and the last added resistance with the value of $(\Gamma_c)^{-1}$ must be in series with all the $(\Gamma_{ij})^{-1}$'s that are *exponentially smaller* than $(\Gamma_c)^{-1}$, owing to the exponential dependence of $\Gamma_{ij}$ on the randomly distributed parameters $E_{ij}$ and $S_{ij}$. It follows that the sum of all the exponentially small resistance would amount to at most a non-exponential multiplicative factor to $\Gamma_c$. Similarly, the conductance $\Gamma_{ij}$ that satisfy $\Gamma_{ij} \leq \Gamma_c$ are in parallel to $\Gamma_c$ (since they offer alternative paths of conduction). But these conductance must be exponentially smaller than $\Gamma_c$, again owing to the exponential dependence of $\Gamma_{ij}$ on $E_{ij}$ and $S_{ij}$. Hence the net effect is that all the other values of $\Gamma_{ij}$ that are *not* in the vicinity of $\Gamma_c$ can at most contribute a constant multiplicative factor that is not too large.

To examine the temperature dependence of $\Gamma_c$, let us consider the equality $\Gamma_{ij} \cong \Gamma_c$, or

$$\Gamma_c \cong \Gamma_o \exp(-E_{ij}/k_B T) \exp(-2\chi S_{ij}). \tag{3}$$

Here $\Gamma_c$ is the conductance of the critical percolating network, determined by the critical conduction condition as specified by Eqs. (4), (5) below. In what follows we show that the temperature dependence of GH is derivable from the CPM with minimal numerical input. The accuracy to which $\Gamma_c$ can describe the conductance of the whole sample may be checked by numerical simulations. Our simulations, based on the immediate neighbor hopping model, yield excellent agreement with the CPM.

By taking the logarithms of both sides of Eq. (3) and using the expression $E_{ij} = \max[E_i, E_j]$, we obtain

$$\frac{S_{ij}}{S_m} + \frac{\max[E_i, E_j]}{E_m} \cong 1, \tag{4}$$

where

$$S_m = \frac{1}{2\chi} \ell n \frac{\Gamma_o}{\Gamma_c}, \tag{5a}$$

$$E_m = k_B T \ln \frac{\Gamma_o}{\Gamma_c}. \tag{5b}$$

In Eq. (4), if $S_{ij}$ and $\max[E_i, E_j]$ are respectively replaced by their values at the critical bond in the percolation cluster, it can be shown through simulations that the equality is indeed very well satisfied.

It is noted that the critical conductance condition as specified by Eqs. (4), (5) can be achieved by an infinite number of combinations of $E_{ij}$ and $S_{ij}$ between the immediate-neighbor pairs, as can be seen from Eq. (3) or (4). As long as $S_{ij}$ and $E_{ij}$ are smaller than their respective $S_m$ and $E_m$, one can have an infinity of possible choices for the two fractions that add up to 1 in Eq. (4). That makes $S_{ij}$ and $E_{ij}$ treatable as independent random variables, with the attainment of the percolation conductance in a random medium a certainty. We elaborate on this point below.

In Eq. (4), if the left hand side is less than 1, then $\Gamma_{ij}$'s are much larger than $\Gamma_c$ (owing to the exponential dependence on the parameters), and the connected clusters do not percolate. However, if the left hand side of Eq. (4) is greater than 1, then the $\Gamma_{ij}$'s are exponentially smaller than $\Gamma_c$. By using the rule of our pair connectivity as stated above, there can be many different percolating paths. For example, if we regard the left hand side of Eq. (4) as the sum of two random numbers each ranging from zero to one, then the probability of having the value (of the sum) around 1 is the highest, since there is a maximum number of combinations, and each random number can fully access the full range of [0,1]. This is in contrast to the probability of having the value around 0 or 2, in which the combination is unique, and each random number on the left hand side must each be limited to be around either 0 or 1. Hence the probability of having such values is much smaller. From this simple discussion it is clear that (1) the immediate-neighbor CPM can treat the energy of activation and the tunneling conduction as two random variables, fully compatible with the requirement of a random medium, and (2) there is a wide range of possible choices for the combination of the two parameters to achieve the critical conductance that can be representative of the sample conductance.

As $S_m$ and $E_m$ both contain $\Gamma_c$ in their definitions, and as temperature $T$ is present in $E_m$, it requires only a relationship between $S_m$ and $E_m$ to deduce the temperature variation of $\Gamma_c$ and hence that of the whole sample. In particular, if it can be shown that $E_m S_m = \text{constant}$, then from Eqs. (5a) and (5b) one can immediately deduce $\Gamma_c \propto \exp\left[-(T_o/T)^{1/2}\right]$, i.e., our goal is achieved. Below we demonstrate that precisely such a relation can be obtained by considering the *average* number of conducting bonds per site $b_c$ (i.e., $\Gamma_{ij} \geq \Gamma_c$, alternatively denoted the average number of bonds per site) on the critical percolating network when the percolation occurs. In our present case the value of $b_c$ ($\approx 1.5$ for the 3D simple cubic lattice [40]) should be that when only the immediate-neighbor hopping is considered.

## III. Critical path derivation of the temperature dependence

*Model system*

To proceed further, we model our granular system by grains arranged in a (locally) simple cubic geometry, with 6 neighbors. The conducting grains are roughly round in shape and can have a distribution of diameters. However, the width of the size distribution should be relatively small. The shape can also deviate from a perfect sphere, so that the closest separation between the surfaces of the neighbors (i.e., the tunneling distance) can have a distribution. Below we calculate the *average* parameters of the system by assuming spherical grains.

If the grains are spherical with an average diameter $d$, and if the average distance of closest approach is $S$, then the volume fraction of conducting grains is given by the volume of the grain, $(\pi/6)d^3$, divided by the volume of the simple cubic unit cell, $(S+d)^3$, i.e., $x = \dfrac{(\pi/6)d^3}{(S+d)^3} = \dfrac{0.524}{[1+(S/d)]^3}$. Here the *averaged ratio* $S/d = \mu$ is a constant of the sample, given by $\mu = (0.524/x)^{1/3} - 1$. We assume that the distance of closest approach between the surfaces of two immediate neighbors can have a distribution, physically caused by the surface or shape irregularities of the grains.

*Evaluation of $b_c$ by using excitation energy and tunneling distance distribution functions*

The calculation of $b_c$, which again is an average quantity, involves simple counting [39]. That is, if we allow $E \ (= E_i)$ and $S \ (= S_{ij})$ to fluctuate independently as argued above, then the *average* number of bonds emanating from each grain, *for the critical percolation network*, can be evaluated as the product of $Z$, the number of immediate neighbors, with the probability of having a "conducting bond" emanating from a grain in the critical percolation network. The definition of a "conducting bond" is defined in the previous section, i.e., any bond that satisfies the condition $\Gamma_{ij} > \Gamma_c$ within the critical percolation network. The above condition can be mathematically expressed as

$$b_c = Z \int_0^{S_m} dS \int_0^{E^u} dE \left[ D(E) G(S) \right] , \qquad (6)$$

where $Z$ (=6 for the simple cubic geometry) denotes the number of immediate neighbors, $E^u = E_m(1 - S/S_m)$ is the upper bound of the excitation energy in the critical network, as governed by Eq. (3), and $D(E)$ is given by Eq. (2). Here $G(S)$ denotes the distribution of $S$, taken to be a flat distribution with width $2S_o$, i.e., $G(S) = 1/2S_o$ for $0 \leq S \leq 2S_o$, and zero otherwise, and $D(E)$ represents the distribution of the excitation energies for the charge carriers, obtained by taking into account the random potential shifts of grain's Fermi level as described in the introduction section. Equation (6) is noted to differ from the traditional CPM approach of using the density of states $\rho$ to evaluate $b_c$. However, the basic logic remains the same, and the direct use of the distribution functions is statistically accurate for the physical scenario of immediate neighbor granular hopping. We would also like to remark that Eq. (6) cannot be used for the traditional VRH scenario because in that case the number of "neighbors" for hopping conduction is not well-defined, i.e., theoretically $Z \to \infty$; hence the use of $\rho$ for the evaluation of $b_c$, in the VRH context, is a rather unique choice.

*Emergence of the granular hopping conduction's temperature dependence*

To use Eq. (6) in the context of a physical model system (such as the model used in the simulation, presented below), we replace $S_m$ and $E_m$ in Eq. (6) by the parameters $S^{\max}$ and $E^{\max}$ which denote, respectively, the upper bounds of inter-grain tunneling distance and excitation energy of the percolating cluster, i.e., $S^{\max} = \max(S_{ij})$ and $E^{\max} = \max(E_{ij})$. By using $S^{\max}$ and $E^{\max}$, one can avoid the a-priori knowledge of $\Gamma_c$, which requires the solution of the problem. In the simulations. $S^{\max}$ and $E^{\max}$ are shown to be proportional to $S_m$ and $E_m$, respectively, with respective proportionality constants $\beta = \beta_S = \beta_E \cong 0.9$, defined by $\beta_S = \langle S_m/S^{\max} \rangle$ and $\beta_E = \langle E_m / E^{\max} \rangle$, where $S_m$ and $E_m$ are given by Eqs. (4) and (5), respectively, with the conductance $\Gamma_c$ obtained self-consistently from the simulation results.

Let us denote $I(E^u) = \int_0^{E^u} D(E)dE$ and $J(S^{\max}) = \int_0^{S^{\max}} G(S)dS$. It is clear that $J(S^{\max}) = S^{\max} / 2S_o$. In Fig. 1 we plot the typical behavior of $I(E^u)$ as a function of $E^u$. It is found that the function is linear at small values of $E^u$ (which is relevant for low temperatures), but is required to saturate at large values of $E^u$ (which applies to high temperatures). This behavior follows directly from the constancy of $D(E)$ in the low energy regime. It follows that if we confine ourselves to the low temperature regime, then (by using the fact that $S^{\max}$ and $E^{\max}$ are proportional to $S_m$ and $E_m$, respectively)

$$b_c \propto E^{\max} S^{\max} \propto E_m S_m . \tag{7}$$

Since we have already argued in the previous section that the constancy of the product $E_m S_m$ implies the GH conduction behavior, it follows from Eq. (7) that at low temperatures at least, the GH conduction must follow the behavior

$$\Gamma_c = \Gamma_o \exp\left[-(T_o / T)^{1/2}\right] . \tag{8}$$

An analytical expression for $T_o$ may be obtained by evaluating the slope of $I(E^u)$ in the limit of $E^u \to 0$. The result is

$$T_0 = 5.2 \frac{\beta^2 \chi b_c \mu e^2}{k_B \kappa(x)} , \tag{9}$$

which represents an upper bound to the value of $T_o$. Here $\kappa(x) = \varepsilon[1+1/2\mu(x)]$ is the dielectric constant that can vary as a function of composition [1, 9]. The derivation of Eq. (9) is given in the Appendix, in which the value of Z=6 and the width of the lognormal distribution, $\omega_E$ =0.2, are both taken into account in arriving at the numerical constant of 5.2 in Eq. (9).

We note here that the linear behavior of $I(E^u)$ vs. $E^u$, at the lower section of the curve, is independent of the parameter values. *Hence the temperature variation as displayed by Eq. (8) is a direct consequence of the flatness of the distributions D(E) and G(S). As such, the temperature dependence of granular hopping should be very broadly applicable to a wide range of granular systems.* The slope of the linear section, however, is dependent on the value of *x*, as well as on the average number of immediate neighbors. The specific parameter values are given in the caption of Fig. 1.

*Direct numerical solution of Eq. (6) and comparison with the experiment*

If we take into account the accurate behavior of $I(E^u)$, then a numerical solution of Eq. (6) yields an extremely good $1/\sqrt{T}$ behavior over a very broad range of temperatures, with only a slight deviation at higher temperatures. This is shown in Fig. 2 for the same parameter values used in Fig. 1. The higher temperature deviation, due to the saturation behavior of $I(E^u)$, is necessarily in the nature of 1/T thermal activation (which would appear as a quadratic in the $1/\sqrt{T}$ plot). This is not surprising, since at infinite temperature the conduction should be dominated by thermal activation only, and the slope of 1/T is given by the maximum activation energy in the percolation path. But up to 100 K, the $1/\sqrt{T}$ behavior is an extremely good description of GH conduction. In Fig. 2 we also show the experimental data on a sample of $Ag_x(SnO_2)_{1-x}$ with $x=33\%$. It is seen that the comparison between theory and experiment yields extremely good agreement.

Because the linear behavior at the lower section of the $I(E^u)$ vs. $E^u$ is independent of the parameter values, hence the functional form of Eq. (8) is a general consequence of the critical path argument as applied to the immediate neighbor hopping. The value for $T_o$, Eq. (9), however, depends on the composition $x$ as well as on the average number of immediate neighbors, e.g., on whether simple cubic or hexagonal closed-packed geometry is chosen. For the present case, locally simple cubic geometry with $Z=6$ is seen to yield very good agreement with the experimental result.

## IV. Discussion

The physical picture of this derivation is that the GH conduction's temperature dependence arises from the bounds on the percolating conduction paths' tunneling and activation parameters, imposed by the average composition of the sample. It relaxes considerably the microscopic constraint imposed by the previous theories [3, 32, 33]. In particular, if now we substitute Eq. (8) back into Eq. (4) and Eq. (5), then $E_m = k_B \sqrt{T_o T}$ and $S_m = \frac{1}{2\chi}\sqrt{\frac{T_o}{T}}$. That is, the range of allowed activation energy diminishes as the temperature is lowered (i.e., shifts to larger grains), whereas the allowed maximum tunneling distance is increased at the same time. However, one can still have an infinity of possible combinations of both parameter values, as long as their sum satisfies the criterion

$$\frac{2\chi S_{ij}}{\sqrt{T_o/T}} + \frac{E_i}{k_B\sqrt{T_o T}} \cong 1 .  \qquad (10)$$

The fact that the values of $S_m$ and $E_m$ have this type of relationship precisely reflects the overall compositional constancy, as evidenced in the presence of the parameter $\mu$ in the $T_o$ expression. The immediate-neighbor CPM approach is able to translate this very weak constraint into the widely observed $\sigma = \sigma_o \exp\left[-(T_o/T)^{1/2}\right]$ behavior.

## V. Numerical simulations

We have performed simulations on 3D simple cubic resistance network by calculating the conductance of the system by using Kirchhoff's laws through nodal analysis. We will use the parameters of an experimental granular system, $Ag_x(SnO_2)_{1-x}$, which demonstrates excellent granular hopping conduction behavior [9]. The excitation energy of the immediate-neighbor grains is extracted from a log-normal distribution, $N(E_i)$, of the charging capacitance energy, convolved with a random disordered potential as described in Ref. [19]. The resulting distribution is given by Eq. (2). Here $E_i$ is the capacitive charging energy of grain $i$. The most probable value in the log-normal distribution of $E_i$ is taken to be $E_o = 2e^2/\kappa(x)d^{(e)} = 6$ meV, with $\varepsilon = 15$, $\mu = 0.17$ (corresponding to $x = 33\%$, in anticipation of comparison with the experimental data with the same value of $x$), and the width of the distribution is taken to be $\omega_E = 0.2$, which is obtained from the experimentally determined grain size distribution given in Ref. [9]. It is to be noted that $\omega_E = 0.2$ represents a relatively narrow size distribution. Here we have assumed $d \approx d^{(e)}$.

The log-normal distribution is given by [19, 25]

$$N(E_i) = \frac{1}{\sqrt{2\pi}\omega_E E_i} \exp\left[-\frac{\ln^2(E_i/E_0)}{2\omega_E^2}\right]. \qquad (11)$$

It should be noted that the abundance of low energies in the convolved distribution $D(E)$ justifies the lower integration limit in Eq. (6). The inter-grain separation between the immediate neighbors is assumed to obey a flat distribution having a mean value of $S_0 = 14$ Å, with $\chi = 0.09$ Å$^{-1}$, so that conforms with $\mu = 0.17$. It is to be noted that the model used in the present simulation is entirely consistent with the immediate-neighbor CPM model.

By averaging over 20 configurations for a $10 \times 10 \times 10$ simple cubic resistance network, we show in Fig. 2 a plot of simulated $\ell n(\Gamma_o / \Gamma_c)$ as a function of $1/\sqrt{T}$. The results of the numerical simulations are seen to be in excellent agreement with the $\exp\left[-(T_o/T)^{1/2}\right]$ behavior of both the experiment and theoretical results over the temperature range of 6 K to 100 K. From the slope, $T_0$ is found to be 231 K. By using the results obtained from simulations, i.e., $\beta \cong 0.9$ and the parameters $b_c \approx 1.5$, $\varepsilon = 15$, $\mu = 0.17$ and $\chi = 0.09$ Å$^{-1}$ (which can vary as a function of $x$, especially close to the metal-insulator transition) the value of $T_o$ predicted by the exact solution of Eq. (6) is 255 K, which is ~10% higher than the simulation result.

To further check the validity of the immediate-neighbor CPM, we study the temperature dependence of the two model parameters, the maximum charging energy $E^{\max}$ and maximum tunneling distance $S^{\max}$, which were extracted from the maxima of the bonded sites for the percolation cluster and are expected to be proportional to $E_m$ and $S_m$, respectively. From our numerical results, we plot $S^{\max}$ against $1/\sqrt{T}$ and $E^{\max}$ against $\sqrt{T}$ as shown in Fig. 3. They clearly show the expected behavior, with the slopes of 106 Å K$^{1/2}$ and 1.48 meV K$^{-1/2}$, respectively. Together with the proportionality constants $\beta$, $T_o$ can also be determined from the following expressions:

$$S^{\max} = \frac{1}{2\chi\beta_S}\sqrt{\frac{T_o}{T}}, \qquad (12a)$$

$$E^{\max} = \frac{k_B}{\beta_E}\sqrt{TT_o}. \qquad (12b)$$

The value obtained for $T_o$ turns out to be around ~270 K, which is very close to the upper bound value predicted by Eq. (9).

## VI. Conclusions

We have revisited the electronic conduction processes in random granular films in the dielectric regime, where the conducting granules are allowed to possess a distribution of sizes which, in turn, govern the magnitude of the charging energy. By taking into account the random potential that can arise from electrical impurities and interfacial states, and by employing the critical path approach applied to the immediate-neighbor pairs of grains, we show that the well-known fractional exponential temperature behavior of conductivity, $\sigma = \sigma_o \exp\left[-(T_o/T)^{1/2}\right]$, emerges naturally from the loose constraint of the average metal volume fraction of the whole sample. Numerical simulations using realistic material parameters agree well with the immediate-neighbor critical path method, as well as with the measured experimental data. The present approach does not invoke any specific material properties or compositions, e.g., the uniformity of the sample (which would be the case for the effective medium approach) and thus the result should be applicable to a wide variety of granular composites, as the experimental results have demonstrated.

**Acknowledgements**

We thank Ya-nan Wu and Zhi-qing Li for useful discussions. This work was supported by GRF Grant No. 16304314 (P.S.), and by the Taiwan Ministry of Science and Technology through Grant No. MOST 106-2112-M-009-007-MY4 and the MOE SPROUT Project (J.J.L.).

**Appendix**

*Derivation of the expression for $T_0$*

Here we show that the accurate expression of $T_0$ is given by

$$T_0 = \frac{16\beta_E \beta_S b_c E_0 S_0 \chi}{k_B Z} \exp\left(-\frac{\omega_E^2}{2}\right). \tag{A1}$$

in the limit of small excitation energies (i.e., low temperatures).

We start with the expression for the excitation energy distribution

$$D(E) = \int_0^\infty N(E_i) f(E - E_i)\, dE_i, \tag{A2}$$

where

$$N(E_i) = \frac{1}{\sqrt{2\pi}\,\omega_E E_0} e^{-\frac{\ln^2\left(\frac{E_i}{E_0}\right)}{2\omega_E^2}}, \tag{A3a}$$

and

$$f(E) = \frac{1}{2E_i}\Theta(E_i - |E|) \tag{A3b}$$

After some algebras, $D(E)$ can be re-written as

$$D(E) = \int_0^\infty N(E_i) f(E - E_i)\, dE_i = \frac{e^{\frac{\omega_E^2}{2}}}{2E_0\sqrt{\pi}} \int_{\frac{\omega_E}{\sqrt{2}} + \frac{\ln\left(\frac{E}{2E_0}\right)}{\omega_E\sqrt{2}}}^\infty e^{-y^2}\, dy. \tag{A4}$$

In the small $E$ limit,

$$\frac{\omega_E}{\sqrt{2}} + \frac{\ln\left(\frac{E}{2E_0}\right)}{\omega_E\sqrt{2}} \to -\infty$$

In this limit, the integral in $D(E)$ gives $\sqrt{\pi}$, and thus the value of $D(E)$ is

$$\frac{\exp\left(\frac{\omega_E^2}{2}\right)}{2}$$

In particular, for $\omega_E = 0.2$, this value is ~0.51, which agrees with what we obtained numerically.

Since

$$\frac{E_i}{E_m} + \frac{S_{ij}}{S_m} \approx 1, \tag{A5}$$

The upper bound for $S_{ij}$ is

$$S^{upper} = S_m\left(1 - \frac{E}{E_m}\right). \tag{A6}$$

Therefore,

$$\beta^2 b_c = Z \int_0^{E_m} dE \int_0^{(1-\frac{E}{E_m})S_m} dS\, [G(S)\, D(E)]$$

$$= Z \int_0^{E_m} dE \int_0^{(1-\frac{E}{E_m})S_m} \frac{1}{2S_0}\left(\frac{\exp\left(\frac{\omega_E^2}{2}\right)/2}{E_0}\right) dS$$

$$= Z\left(\frac{1}{2S_0}\right)\left(\frac{\exp\left(\frac{\omega_E^2}{2}\right)/2}{E_0}\right)\int_0^{E_m} dE\left(1 - \frac{E}{E_m}\right)S_m$$

$$= Z\left(\frac{1}{2S_0}\right)\left(\frac{\exp\left(\frac{\omega_E^2}{2}\right)/2}{E_0}\right)\left(\frac{1}{2}\right)E_m S_m$$

$$= \frac{Z\left[\exp\left(\frac{\omega_E^2}{2}\right)/2\right]}{4E_0 S_0} E_m S_m$$

$$= \frac{Z\left(exp\left(\frac{\omega_E^2}{2}\right)/2\right)}{4E_0 S_0}\left(\frac{k_B T_0}{2\chi}\right). \tag{A7}$$

From Eq. (A7) we obtain

$$T_o = \frac{16\beta^2 b_c E_o S_o \chi}{k_B Z}\exp\left(-\frac{\omega_E^2}{2}\right). \tag{A8}$$

Furthermore, by using $E_0 = 2e^2/\kappa d$ and $\mu = S_0/d$, we have

$$T_o = \frac{16\beta^2 b_c e^2 \mu \chi}{k_B \kappa Z\left(\frac{1}{2}\right)\exp\left(\frac{\omega_E^2}{2}\right)}. \tag{A9}$$

Since $exp\left(\frac{\omega_E^2}{2}\right)/2 \approx 0.51$ and $Z = 6$, the numerical coefficient becomes $16/Z(0.51) = 5.2$, hence

$$T_0 = 5.2\frac{\beta^2 b_c e^2 \mu \chi}{k_B \kappa(x)}$$

which is Eq. (9) in the manuscript.

## References

[1] Abeles B, Sheng P, Coutts M D and Arie Y 1975 Structural and electrical properties of granular metal films *Advances in Physics* **24** 407-61

[2] Adkins C J 1989 Conduction in granular metals--variable-range hopping in a Coulomb gap? *Journal of Physics: Condensed Matter* **1** 1253

[3] Sheng P, Abeles B and Arie Y 1973 Hopping conductivity in granular metals *Physical Review Letters* **31** 44-7

[4] Barzilai S, Goldstein Y, Balberg I and Helman J S 1981 Magnetic and transport properties of granular cobalt films *Physical Review B* **23** 1809-17

[5] McAlister S P, Inglis A D and Kayll P M 1985 Conduction in cosputtered Au-SiO$_2$ films *Physical Review B* **31** 5113-20

[6] McAlister S P, Inglis A D and Kroeker D R 1984 Crossover between hopping and tunnelling conduction in Au-SiO$_2$ films *Journal of Physics C: Solid State Physics* **17** L751

[7]   Bakkali H and Dominguez M 2013 Differential conductance of Pd-$ZrO_2$ thin granular films prepared by RF magnetron sputtering *EPL (Europhysics Letters)* **104** 17007
[8]   Gantmakher V F 2005 *Electrons and Disorder in Solids* (Oxford: Clarendon)
[9]   Wu Y N, Wei Y F, Li Z Q and Lin J J Granular hopping conduction in $(Ag,Mo)_x(SnO2)_{1-x}$ films in the dielectric regime *arXiv:1708.04434*
[10]  Mott N F 1968 Conduction in glasses containing transition metal ions *J. Non-crystal. Solids* **1** 1
[11]  Mott N F and Davis E A 1979 *Electronic Processes in Non-Crystalline Materials* (Oxford: Clarendon)
[12]  Efros A L and Shklovskii B I 1975 Coulomb gap and low temperature conductivity of disordered systems *Journal of Physics C: Solid State Physics* **8** L49
[13]  Efros A L 1976 Coulomb gap in disordered systems *Journal of Physics C: Solid State Physics* **9** 2021
[14]  Shklovskii B I and Efros A L 1984 *Electronic Properties of Doped Semiconductors* (New York: Springer)
[15]  Miller A and Abrahams E 1960 Impurity Conduction at Low Concentrations *Physical Review* **120** 745-55
[16]  Ambegaokar V, Halperin B I and Langer J S 1971 Hopping Conductivity in Disordered Systems *Physical Review B* **4** 2612-20
[17]  Shante V K S 1973 Variable-range hopping conduction in thin films *Physics Letters A* **43** 249
[18]  Shante V K S 1977 Hopping conduction in quasi-one-dimensional disordered compounds *Physical Review B* **16** 2597-612
[19]  Adkins C J 1987 Conduction in granular metals with potential disorder *Journal of Physics C: Solid State Physics* **20** 235
[20]  Pollak M and Adkins C J 1992 Conduction in granular metals *Philosophical Magazine B* **65** 855-60
[21]  Adkins C J, Benjamin J D, Thomas J M D, Gardner J W and McGeown A J 1984 Potential disorder in granular metal systems: the field effect in discontinuous metal films *Journal of Physics C: Solid State Physics* **17** 4633
[22]  McGeown A J and Adkins C J 1986 Thermopower in discontinuous metal films *Journal of Physics C: Solid State Physics* **19** 1753
[23]  Cavicchi R E and Silsbee R H 1984 Coulomb Suppression of Tunneling Rate from Small Metal Particles *Physical Review Letters* **52** 1453-6
[24]  Gardner J W and Adkins C J 1985 Island charging energies and random potentials in discontinuous metal films *Journal of Physics C: Solid State Physics* **18** 6523
[25]  Buhrman R A and Granqvist C G 1976 Log-normal size distributions from magnetization measurements on small superconducting Al particles *Journal of Applied Physics* **47** 2220-2
[26]  Zhang J and Shklovskii B I 2004 Density of states and conductivity of a granular metal or an array of quantum dots *Physical Review B* **70** 115317
[27]  Beloborodov I S, Lopatin A V and Vinokur V M 2005 Coulomb effects and hopping transport in granular metals *Physical Review B* **72** 125121
[28]  Lin C H and Wu G Y 2000 Hopping conduction in granular metals *Physica B: Condensed Matter* **279** 341-6
[29]  Efetov K B and Tschersich A 2003 Coulomb effects in granular materials at not very low temperatures *Physical Review B* **67** 174205
[30]  Chui T, Deutscher G, Lindenfeld P and McLean W L 1981 Conduction in granular aluminum near the metal-insulator transition *Physical Review B* **23** 6172-5
[31]  Lin Y H, Sun Y C, Jian W B, Chang H M, Huang Y S and Lin J J 2008 Electrical transport studies of individual $IrO_2$ nanorods and their nanorod contacts *Nanotechnology* **19** 045711
[32]  Sheng P and Klafter J 1983 Hopping conductivity in granular disordered systems *Physical Review B* **27** 2583-6

[33]    Klafter J and Sheng P 1984 The Coulomb quasigap and the metal-insulator transition in granular systems *Journal of Physics C: Solid State Physics* **17** L93

[34]    Adkins C J 1990 *Hopping and Related Phenomena,* ed H Fritzsche and M Pollak (Singapore: World Scientific) pp 93-109

[35]    Zhang H, Lu J, Shi W, Wang Z, Zhang T, Sun M, Zheng Y, Chen Q, Wang N, Lin J-J and Sheng P 2013 Large-scale Mesoscopic Transport in Nanostructured Graphene *Physical Review Letters* **110** 066805

[36]    Huang Y-L, Chiu S-P, Zhu Z-X, Li Z-Q and Lin J-J 2010 Variable-range-hopping conduction processes in oxygen deficient polycrystalline ZnO films *Journal of Applied Physics* **107** 063715

[37]    M.V. Feigelman, A.S. Ioselevich 2005, Variable-range cotunneling and conductivity of a granular metal, *JETP Letters*, 81, 277

[38]    C -H Lin and G Y Wu 2001 Percolation calculation with non-nearest neighbor hopping of hopping resistances for granular metals  *Thin Solid Films* **397**, 280

[39]    Sheng P 2006 *Introduction to Wave Scattering, Localization and Mesoscopic Phenomena* (Heidelberg: Springer)

[40]    G E Pike and C H Seager 1974 Percolation and conductivity: A computer study. I *Physical Review B* **10**, 1421

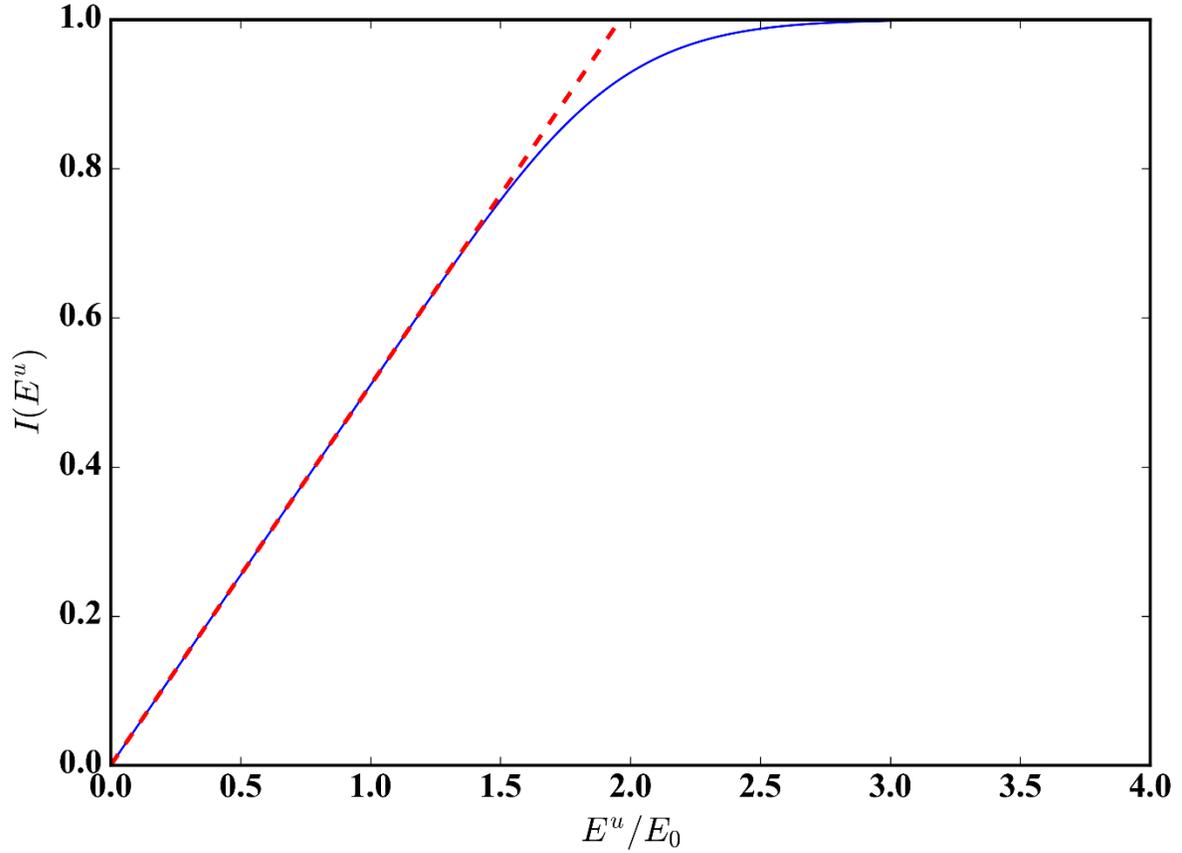

**Fig. 1.** Plot of $I(E^u)$ against $E^u/E_o$. The width of the log-normal distribution $N(E)$, $\omega_E$, is taken to be 0.2 so as to be in agreement with the experimentally measured grain size distribution as given in Ref. [9]. The red dashed line shows the linear regime of $I(E^u)$ at small values of $E^u/E_o$. It can be seen that $I(E^u)$ begins to saturate at $E^u/E_o \geq 2$, contributing to the transition from the $1/\sqrt{T}$ behavior to the thermal activation ($1/T$) behavior of $\ell n(\Gamma_o/\Gamma_c)$ shown in Fig. 2. While the linear behavior of $I(E^u)$ vs. $E^u/E_o$ at the lower section of the curve is perfectly general, whereas the slope of the linear section is dependent on the values of the composition $x$, as well as the number of immediate neighbors. In the text it is shown that the linear section of the curve is responsible for the granular hopping conduction's widely-observed temperature dependence. Hence the functional behavior of Eq. (8) is general; it is a direct consequence of the critical path argument as applied to the immediate neighbor hopping model. For the slope of the linear section of the curve shown we have used $\varepsilon = 15$, $\mu = 0.17$, $\chi = 0.09$ Å$^{-1}$, $\beta = 0.9$ and $b_c = 1.5$. The value of $\mu = 0.17$ is noted to be consistent with $x=33\%$, which is the composition value of the Ag$_x$(SnO$_2$)$_{1-x}$ sample whose temperature dependence of conductance is compared with the theory prediction in Fig. 2.

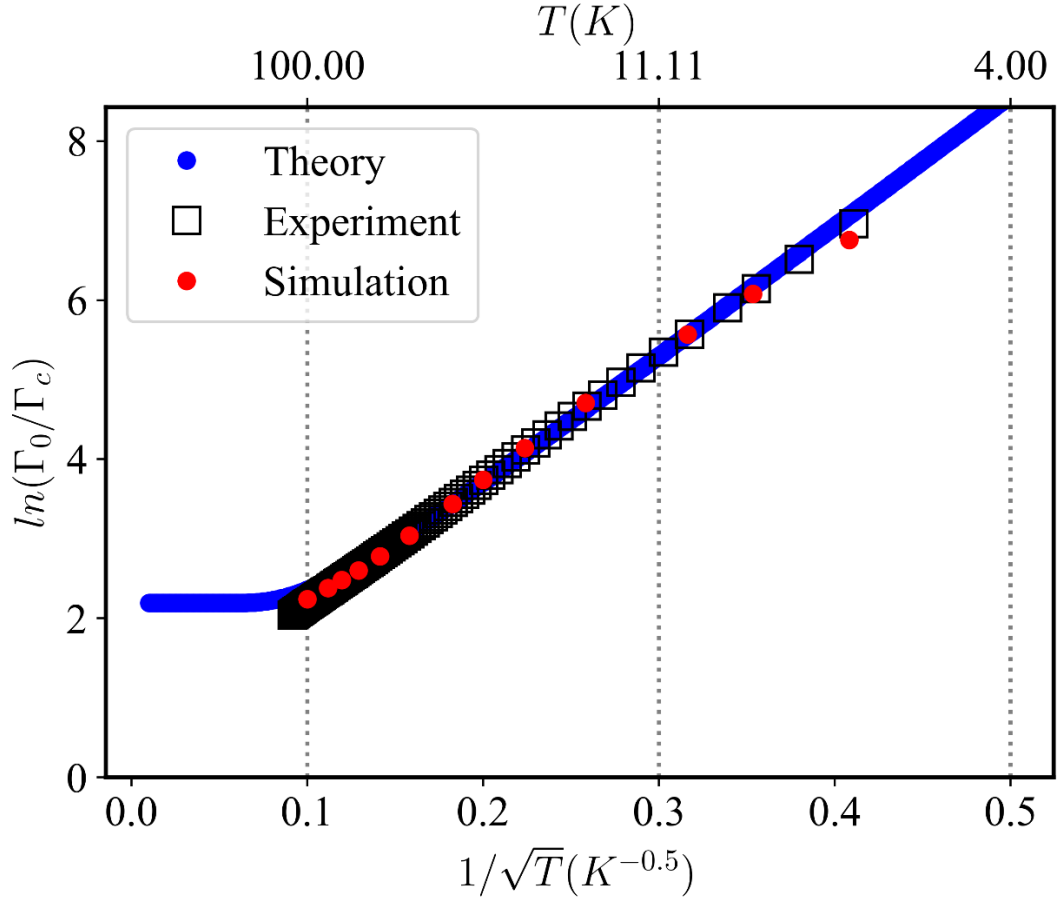

**Fig. 2.** Plots of the logarithm of resistance versus $1/\sqrt{T}$. The theory curve (blue dots) is obtained by solving Eq. (6) for $ln(\Gamma_o/\Gamma_c)$ numerically with the same parameter values as those used in Fig. 1, i.e., $\varepsilon = 15$, $\mu = 0.17$, $\chi = 0.09$ Å$^{-1}$, $\beta = 0.9$ and $b_c = 1.5$. The composition value $x=33\%$, used in the theory and simulation, is that determined experimentally for the Ag$_x$(SnO$_2$)$_{1-x}$ sample. The linear regime of $ln(\Gamma_o/\Gamma_c)$ corresponds to the granular hopping behavior at low temperatures while the quadratic regime of the curve corresponds to thermal activation behavior at high temperatures. The experiment curve (black squares) are reproduced from the experimental data on Ag$_x$(SnO$_2$)$_{1-x}$ (Ref. [9]). The simulation curve (red dots) is obtained by numerically evaluating the conductance of the system based on Kirchhoff laws on a $10\times10\times10$ simple cubic resistance network, averaged over 20 configurations. By using a least square fit, the simulation result for $T_0$ is found to be 231 K. The solution to Eq. (6), for temperatures below 100 K, yields a $T_o = 255$ K. It is noted that the agreement between theory and experiment is extremely good.

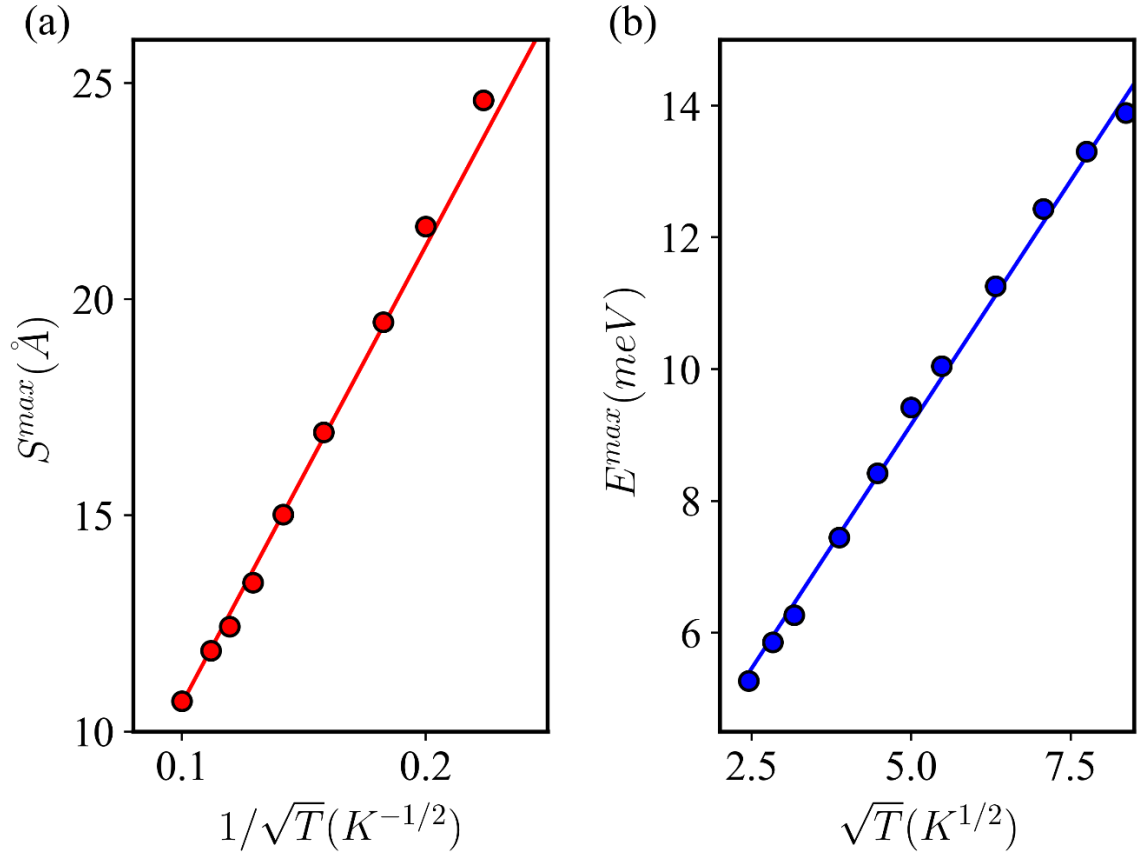

**Fig. 3.** Additional verification of the CPM predictions. **(a)** The grain separation $S^{max}$ on the critical network plotted as a function of $1/\sqrt{T}$, and **(b)** the excitation energy of the critical percolating sites plotted as a function of $\sqrt{T}$. The solid lines are the least-squares fits to the plots. They display the predicted temperature dependence as derived from the CPM approach. The $T_o$ value obtained from the slopes is around 270 K, close to the upper bound prediction of Eq. (9) by using parameters same as those for Fig. 2.